# A SIMPLE AND RELIABLE TOUCH SENSITIVE SECURITY SYSTEM

Adamu Murtala Zungeru[1], Jonathan Gana Kolo[2] and Ijarotimi Olumide[3]

[1,2]School of Electrical and Electronic Engineering, University of Nottingham, Jalan Broga, 43500 Semenyih, Selangor Darul Ehsan, Malaysia
keyx1mzd@nottingham.edu.my, jgkolo@gmail.com
[3]Electrical and Electronic Engineering Technology, Rufus Giwa Polytechnic, Owo, Ondo, Nigeria
ijarotimiolumide@yahoo.com

## ABSTRACT

*This research focuses on detection of unauthorized access to residential and commercial buildings when the residents are far away from the access gate of the house. The system is a simple and reliable touch activated security system and uses sensor technology to revolutionize the standards of living. The system provides a best solution to most of the problems faced by house owners in their daily life. Due to its simple electronic components nature, it is more adaptable and cost-effective. The system is divided into three units; the power supply unit which employs the use of both DC battery and mains supply to ensure constant power supply to the circuit, the trigger unit which is responsible for activating the alarm unit and designed to have much time and period and moderate sensitivity in order to reduce the rate of false alarm, and the alarm amplitude unit which main function is to produce amplitude alarm sound when triggered by the trigger unit with the aim of producing a large audible sound that can alert the entire neighborhood or scare an intruder away. The design of the system was achieved by considering some factors such as economy, availability of components and research materials, efficiency, compatibility and portability and also durability in the design process. The performance of the system after test met design specifications. This system works on the principle of touch sensor. The general operation of the system and performance is dependent on the presence of an intruder entering through the door and touching any part of the door. The overall system was constructed and tested and it work perfectly.*

## KEYWORDS

*Touch Sensor, Automation, Control System, Security, Electronic Circuit, Maximum Power Transfer*

## 1. INTRODUCTION

Insecurity and crime constitute some of the major problems facing our immediate society today. People live with fear of being attacked by burglars, vandals and thieves. Despite all the effort, resources and time that has been devoted to the development of tools that will reduce crime rates and make the world a safer place to live, these problems are still on the increase. These gave rise to the need for an increasing development in the technology of alarm systems which utilizes various principles such as infrared motion detection, light (photo) sensitive electronic devices and so on. Even with the introduction of these alarm systems which have reduced greatly the level of insecurity, there is still a problem of false alarm which needs to be minimized [1]. In order to effectively reduce the level of insecurity and avoid false alarms which can create unnecessary unrest, a touch activated security system is required. This system if properly designed will provide security and ensure alarms are activated only when an unauthorized person try to gain access to the protected area or device by touching the entrance or any other part of the device.

An alarm is a loud noise or signal for alerting or informing people of danger or a problem. An alarm system is thus a security system that produces a form of sound to warn people of a particular danger. The development of alarm systems started with the creation of man. Man required giving alert information and adopted a form of signaling, exclamation and shouting. This was later replaced by clapping of hands and beating of gongs by town criers to alert the community in order to disseminate information in the early African society. All these methods of raising alert were crude, unreliable and inefficient.

With the advancement in science and technology, these crude methods of generating alarm were replaced by electronic alarm systems in the late eighteenth century [2]. These electronic alarm systems operate without any human effort. Once it senses a particular signal, it gives an indication in form of loud sound or noise depending on its design [3]. The first electronic fire alarm system was developed by Dr. William f. Channing and constructed by Mr.-Moses G. Farmer, an electrical engineer [4]. This system uses boxes with automated signaling to indicate the location of fire and was first put into operation in Boston in the United State of America (U.S.A). The development of this alarm system by Dr William was followed with the evolution of fire and burglar alarm technology of varying complexities and sophistications which are too numerous to mention. Notable among these technologies is the remote signaling intruder alarm system which was invented in the early 1970's [2]. This provided a rapid art full response to alarm calls.

However, industries based on security service equipment provision have being coming up with different designs so as to keep burglars and vandals away from public areas not made for them [5]. Today we have the new generation of electronic alarm system which comes in various levels of complexities and sophistication [6]. With the recent increase in crime rates, it has become important to protect our buildings and properties with adequate safety devices with increased level of sophistication [5]. The cost of these safety devices depend on the equipment technology and the application requirement. These safety devices are called the modern electronic alarm system [7]. Some of the modern alarm systems commonly used these days is burglar alarms, duress alarms, industrial alarms, speed limit alarms, and anti-theft car alarms. The burglar alarm is made from a complete electronic circuit loop where by the loop is closed with a bell at the output or a siren so as to alert the owner of what so ever is to be secured. A central control box monitors several motion detectors and perimeter guards and sound an alarm when any of them are triggered. Some burglar alarms work on the concept of magnetic contacts and others on sensitivity. For those that work on sensors, the sensors are normally placed in the entrance of any of the building or restricted entrance of the rooms, in which if the sensor receive a signal above the threshold value set for it, it activate an alarm[7]. In the case of motion detection, ultrasonic method is normally utilized, whereas point detector burglar alarm helps in the detection of theft or unauthorized person at a particular point or a location such as doors or windows. For example, when a particular area is to be monitored, area detection of intruders is utilized within the protected space or location, this is with the help of ultrasonic transducers and/or passive infrared sensors (IR), and are normally mounted at an appropriate location.

Nowadays, closed circuit televisions are incorporated to burglar alarms to detect the presence of unauthorized persons. The output of burglar alarm systems can range from siren or loud alms to telephone automatic dialers and flashing outdoor lights. This serves the function of alerting the neighbors of possible intrusion and also serves as a signal to the police [8]. Auto dialers attached to burglar alarms are programmed to dial the police and play a pre recorded message that informs the police of the house being burgled.

The increasing developments in science and technology have given rise to a tremendous improvement in the technology of alarm sensors. These sensors act as inputs which triggers the alarm. Some of the alarm sensor technologies that have evolved over the years are (1) Microwave sensors: These are motion detection devices that flood a designated area or zone with an electronic field. A movement in the zone or area disturbs the speed and sets off the alarm [9]. (2) Vibration Sensors: The vibration sensors are usually mounted on walls, ceiling and floor with the intention of detecting mechanical vibrations caused by chopping, drilling or any type of physical intrusion [9]. (3) Photo electric beam sensor: These types of sensors transmit infrared rays in the form of light beam to a receiver which is normally in a remote area, hence creating an electronic fence. These sensors are often used to cover openings such as doorways or hall ways acting essentially as trip wire. Once the beam is broken or interrupted, an alarm signal is generated [9]. (4) Electrical field sensors: These sensors generate an electrostatic field between and around an array of conductors and an electrical ground. They detect changes or distortion in the field. This can be caused by anyone approaching or touching the sensor [9]. (5) Audio sensors: These sensors respond to noise generated by intruders entry into a protected area and are generally used but exclusively for internal application. (6) Capacitance sensors: These sensors detect changes in electric field. When a signal is generated, which is either below or higher in value than the threshold signal level when an intruder happens to effect changes in the capacitance of the field, which is as a result of the intruder getting closer to or by direct contact with the sensor wire [9].

The alarm sensor technology employed in this system is the capacitance sensor. The sensor's wire is usually attached to metallic door handles, metallic, and protectors etc. It normally require touch or close proximity and subsequent alteration of capacitance to trigger the alarm. This sensor type adopted in this system helps in the reduction of false triggering of alarms and the detection of intruders is not in any way affected by weather condition, radio frequency or electromagnetic disturbance.

The problem of frequent power outages in some part of the world today gave rise to the use of an automatic power change over switch which has the ability of switching between the mains supply and DC battery using a relay. This provides a passive (or standby) redundancy for the power supply to the alarm system thereby increasing its reliability. The system has both security application and luxury, since it is more comfortable and easy for an intruder to be automatically detected without the need to employ a security guard.

This article is outline in to five major sections. The First Section gives a general introduction of a simple and reliable touch sensitive security system. Section two gives presentation of related works to the aim of this research. In Section Three, detail descriptions of the design and implementation procedures are presented. Section Four presents the experimental results and discussion of the results. In Section Five, we conclude the work with some recommendations. Finally, references used in the manuscript are presented at the end of the paper.

## 2. RELATED WORK

The idea of using sensing devices for convenience, safety, security and quality of service purpose is not new, but the application, cost, design method and reliability of the system varies. In [10,11], the authors consider the use of infra-red rays to count the number of passengers in a car and also remotely control home appliances via short message services for the purpose of security and human convenience. In [3, 5], [8], a burglar alarm system was designed. The burglar alarm is made from a complete electronic circuit loop where by the loop is closed with a bell at the output or a siren so as to alert the owner of what so ever is to be secured. A central control box monitors several motion detectors and perimeter guards and sound an alarm when any of them are triggered. Some burglar alarms work on the concept of magnetic contacts and others on sensitivity. For those that work on sensors, the sensors are normally placed in the entrance of any of the building or restricted entrance of the rooms, in which if the sensor receive a signal above the threshold value set for it, it activate an alarm.

Nowadays, closed circuit televisions are incorporated to burglar alarms to detect the presence of unauthorized persons. The output of which is normally a siren or loud speaker alarms to telephone automatic dialers or lighting systems. This serves the function of alerting the neighbors of possible intrusion and also serves as a signal to the police [8]. Auto dialers attached to burglar alarms are programmed to dial the police and play a pre recorded message that informs the police of the house being burgled. In a related work [12], a Modern duress alarm was considered, and is generally electronic devices that vary widely in capabilities. They are used when under threat to send alarm signals to specific location, and are normally categorized as duress alarm systems. The types of duress alarm systems are:

(i)     Identification alarm: In this system, a simple portable device is used to identify the access able personnel (owner) of the device.

(ii)    Panic button alarm. A push button mounted in a fixed location.

(iii)   A location alarm: a portable device that locates and tracks the person who activated alarm.

The panic button is the most common type of duress alarm. It is found in schools, banks, offices etc. A pre recorded message specifying the location and urgency is sent to several locations such as police department or other security agencies whenever there is a trigger of the system. When the panic button is pushed, a wireless signal is transmitted to the nearest installed wireless repeater unit, which sends the signal to an alarm console. This system does not give specific locations other than the general pre programmed zone of the repeater. The location alarm is a similar version of the identification. The electronics and software of such a system produces a positioning symbol on a console panel or map like display.

Besides, most of the papers mentioned above does not consider cost, reliability and durability in their design procedure, and above all, this paper uses simple and easy to get components to achieve its desired goals.

## 3. SYSTEM DESIGN AND IMPLEMENTATION

This section will discuss the design procedure and the basic theory of components used for this work. The section is further divided into two sub-sections as design theory and system design analysis.

### 3.1. Stages of Implementation

The system design was implemented in three units as shown in Fig. 1. These units are:
(A)     The power supply unit
(B)     The trigger unit
(C)     The alarm/amplifier units

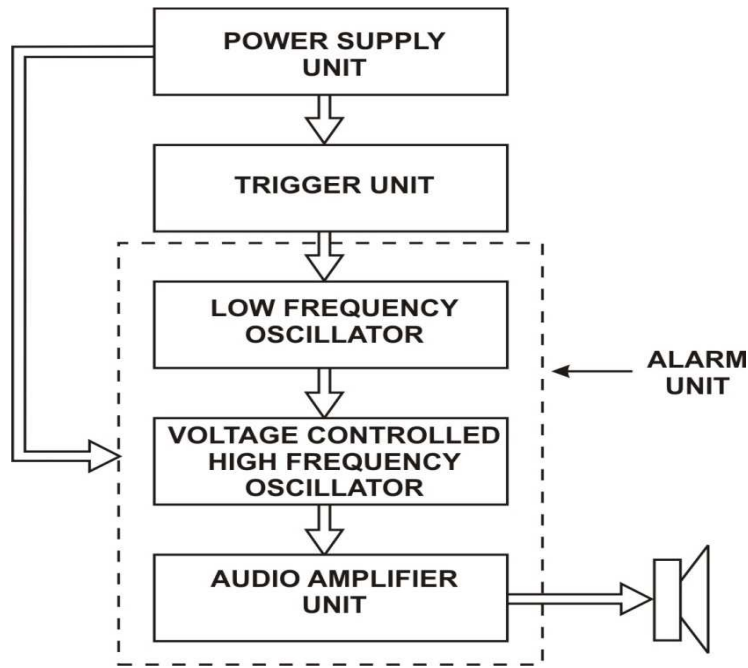

Fig. 1: Block diagram of a touch activated security system

### 3.1.1. Power Supply Unit

The power supply unit is a 2-way automatic power supply system. It gets input from both mains supply and battery supply. The two independent supply systems are connected to a relay switch which acts as an automatic change over switch to switch on any of the available input supply to the main circuit. The power supply unit provides power supply to the other two units of the circuit.

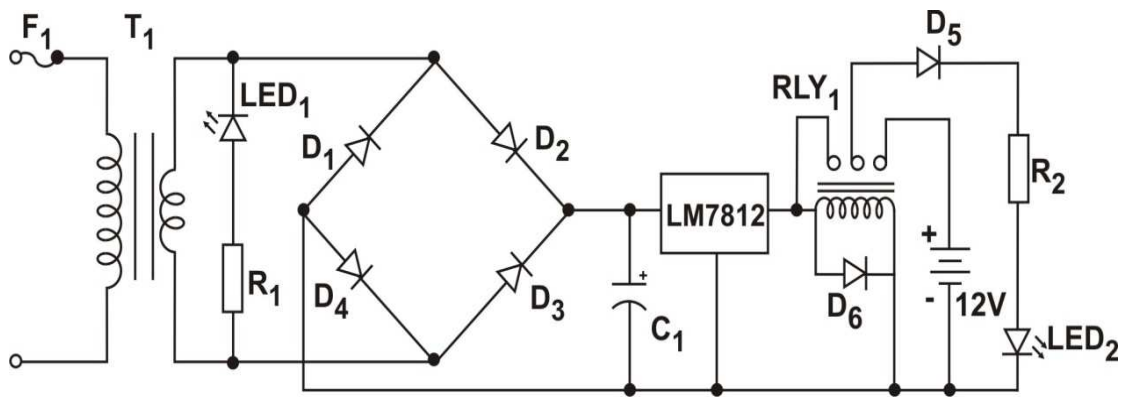

Fig. 2: The schematic diagram of the power supply unit is shown below.

The schematic diagram of the power supply unit consists of two supply input sources (i.e. mains supply and 12V D.C supply from battery). F1 is a protective fuse used to prevent excess current from entering the circuit. T1 is a step down transformer. D1, D2, D3, and D4 are rectifier diodes. C1 is a filter capacitor. IC1 is a regulator IC. Rly1 is a relay switch. R1 and R2 are current limiting resistors protecting LED1 and LED2 respectively.

LED1 is used to indicate the presence of mains supply while LED2 is used to indicate that current is entering the trigger unit. D5 and D6 are protective diodes.

### 3.1.1.1. Operation of the Power Supply Unit

The operation of the power supply unit can be illustrated by the block diagram shown in Fig. 3 below.

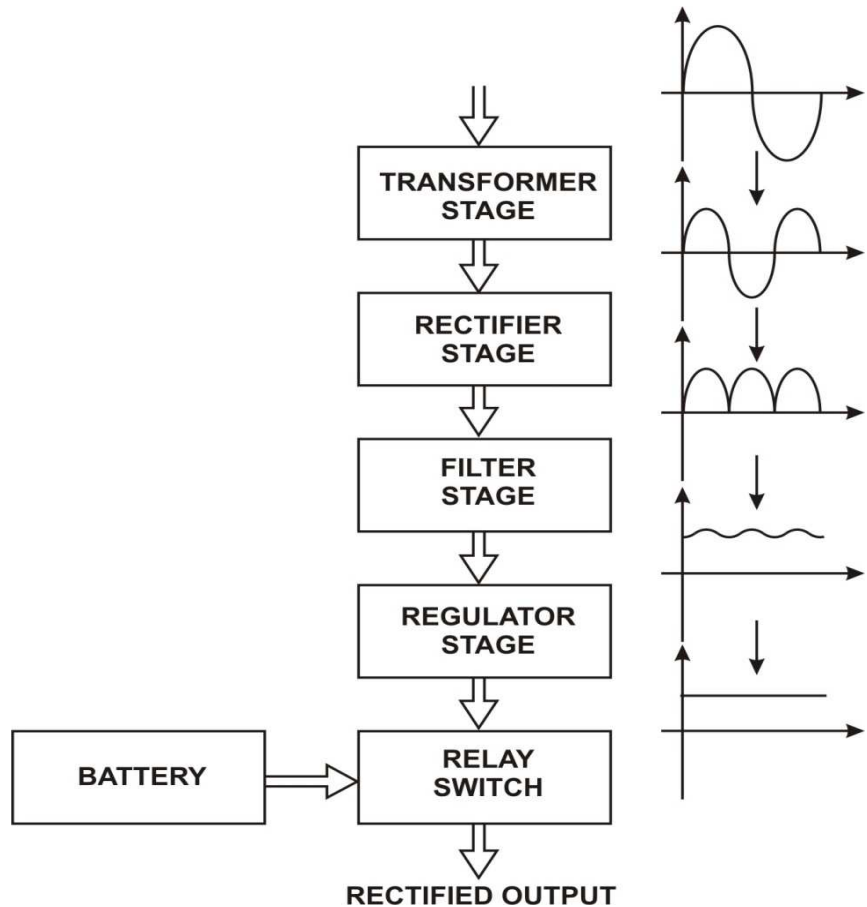

Fig.3: Block diagram of the power supply unit

The block diagram consist of 4 stages for rectification of 240V (A.C) mains supply to 12V (D.C), a battery supply and a relay switch. The description of each stage is given below:

### 3.1.1.2. Transformer Stage

This stage consists of a 240V/18V, step down transformer. It converts the 240V (A.C) voltage supply from mains to 18V (A.C), a 1A fuse (F1) was incorporated at the primary side of the transformer to protect it from excess current. The 18V (A.C) supply is then passed to the rectifier stage. A 220V/18V step down transformer was chosen because the regulator used required more than 12V for its operation.

### 3.1.1.3. Rectifier Stage

In this stage, the rectifier converts the 18V (A.C) supply from the transformer into a pulsating D.C voltage. A full bridge rectifier was used for this purpose. It consist of four diodes (IN 4001 series) arranged as shown in Fig. 2. During the positive half cycles diodes D2 and D3 are forward biased and current flows through the terminals. In the negative half cycle, diodes D1 and D4 are forward biased. Since load current is in the same direction in both half cycles, full wave rectifier signal appears across the terminals [13].

### 3.1.1.4. Filter Stage

The pulsating D.C voltage that comes out from the rectifier stage is converted into constant D.C voltage with the aid of a filter capacitor (C1). This capacitor is large value electrolytic capacitor. It charges up (i.e. store energy) during the conduction half cycle thereby opposing any changes in voltage. The filter stage therefore filters out voltage pulsations (or ripple).

### 3.1.1.5. Regulator Stage

The output of the filter stage varies slightly when the load current or output voltage varies and it is an 18V D.C supply which is higher than the circuit requirement. For these reasons, an LM 7312 Regulator was used to stabilize the voltage and also reduce it from 18V to a 12V steady D.C supply.

### 3.1.2. The Trigger Unit

Below is the schematic diagram of the trigger unit:

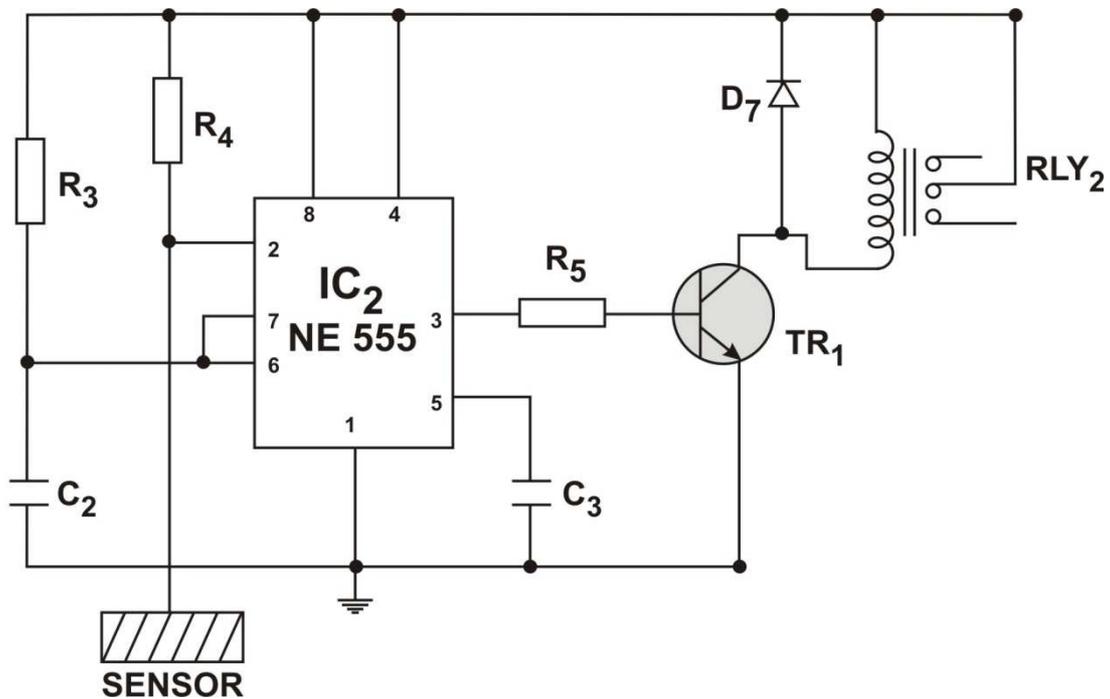

Fig. 4: Circuit diagram of the trigger unit

As shown in Fig. 4, the schematic diagram of the trigger unit consist of 3 major components which areNE555 timer (IC2), Transistor (TR1) and a relay (RLY2). The 555 timer (IC2) produces a trigger current which comes out through its pin 3 whenever pin 2 is activated through the sensor. Pins 4 and 8 are connected to positive power supply while pin 1 is grounded. R3 and C2 determines the time out period of the 555 timer (i.e. the period at which the alarm sound) while R4 determines the sensitivity of the sensor. The output from pin3 (trigger current) is amplified by transistor (Tr1). R5 act as base resistor to Tr1 which is operating in common emitter mode. The output current from transistor (Tr1) causes the relay (Rly2) to operate thereby switching on the alarm/amplifier unit to power supply for duration of time determined by the time out period of the 555 timer (IC2). D7 acts as a commutation diode protecting the transistor (Tr1) from Back-EMF generated by the relay coils. The 555 timer in this unit operates in a mono stable mode.

### 3.1.3. The Alarm Unit

The schematic diagram of the alarm unit is shown in the Fig. 5 below:

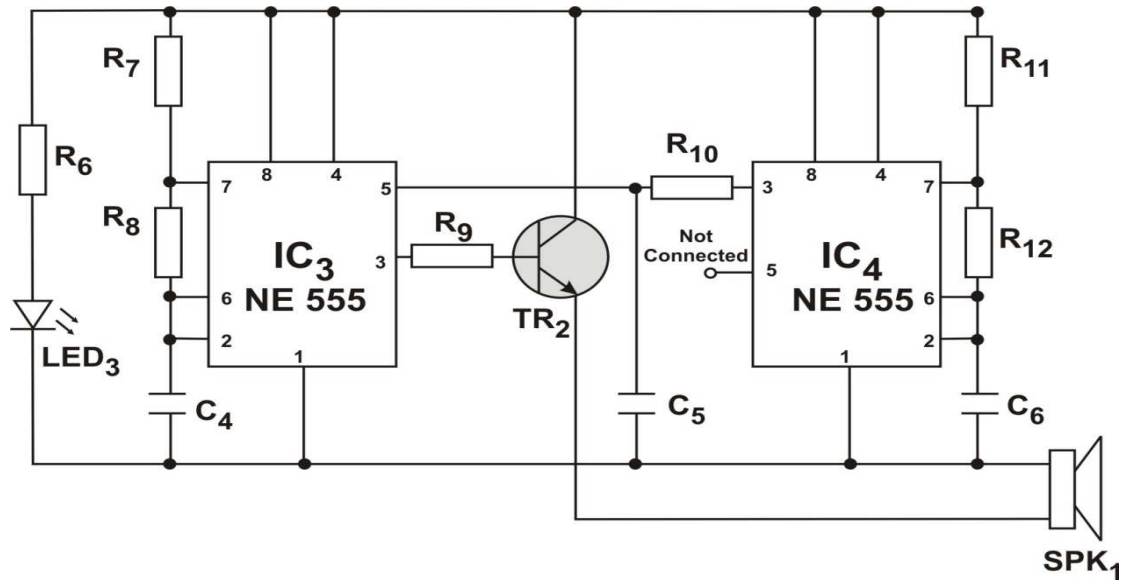

Fig. 5: Circuit diagram of the alarm unit

The alarm unit, as shown in the Fig. 5 above consists of 3 basic components which are; Two 555 timers IC3 and IC4 operating in a stable mode to produce a sire sound and a power transistor (Tr2) used for further amplification of the audio output. IC3 operates at a high frequency of about 481Hz, and act as a voltage controlled oscillator, while producing a square wave. This forms the basic tone of the siren sound system. IC4 produces another square wave much lower frequency of about (0.5Hz) this lower frequency alters the rhythm of the steady tone from IC3 to the desired siren sound. The output of IC4 (i.e. it's pin3) is actually coupled through R9 to control the voltage terminal of IC3. The low frequency (0.5Hz) output from IC4 is used to modulate the high frequency (481Hz) produced by IC3 thereby alternating the frequency of operation of IC3 to produce a siren sound instead of a continuous 481Hz tone.

The final siren note is available at pin3 of IC3 but its maximum current (as calculated on Section 3.3.3) is 0.038A. This current is not sufficient for 5w, 8ohms speaker. The pin3 output of IC3 is therefore fed to the transistor Tr2 for further amplification enabling it to power the speaker thereby producing a very loud audible siren sound.

### 3.2. General Circuit Operation

The Complete circuit diagram of the alarm system is shown in Fig 6 and a table listing the components used (Table 1) is also shown below. The 555 timer in the trigger unit gets activated whenever pin 2 senses a smaller potential that is less than 1/3 the supply voltage. When activated it sounds for duration of time determined by R3 and C2, this also determines how long the alarm will sound before going off. In reality the sensor wire is usually connected to metallic door handles, protectors (for windows and doors) or even metallic gates. Whenever the human body comes in contact with such materials the trigger unit gets activated and the alarm sounds. A detailed explanation of this action is given by the following points. Pin 2 actually acts as a capacitance sensor and detects changes in electrostatic fields. The human body and most conductors in nature generate electrostatic charges which alters the capacitance of the field thereby inducing a small signal on the sensor wire. This signal is less than 1/3 the supply voltage thereby causing the trigger unit to operate for a period of time determined by C2 and R3.

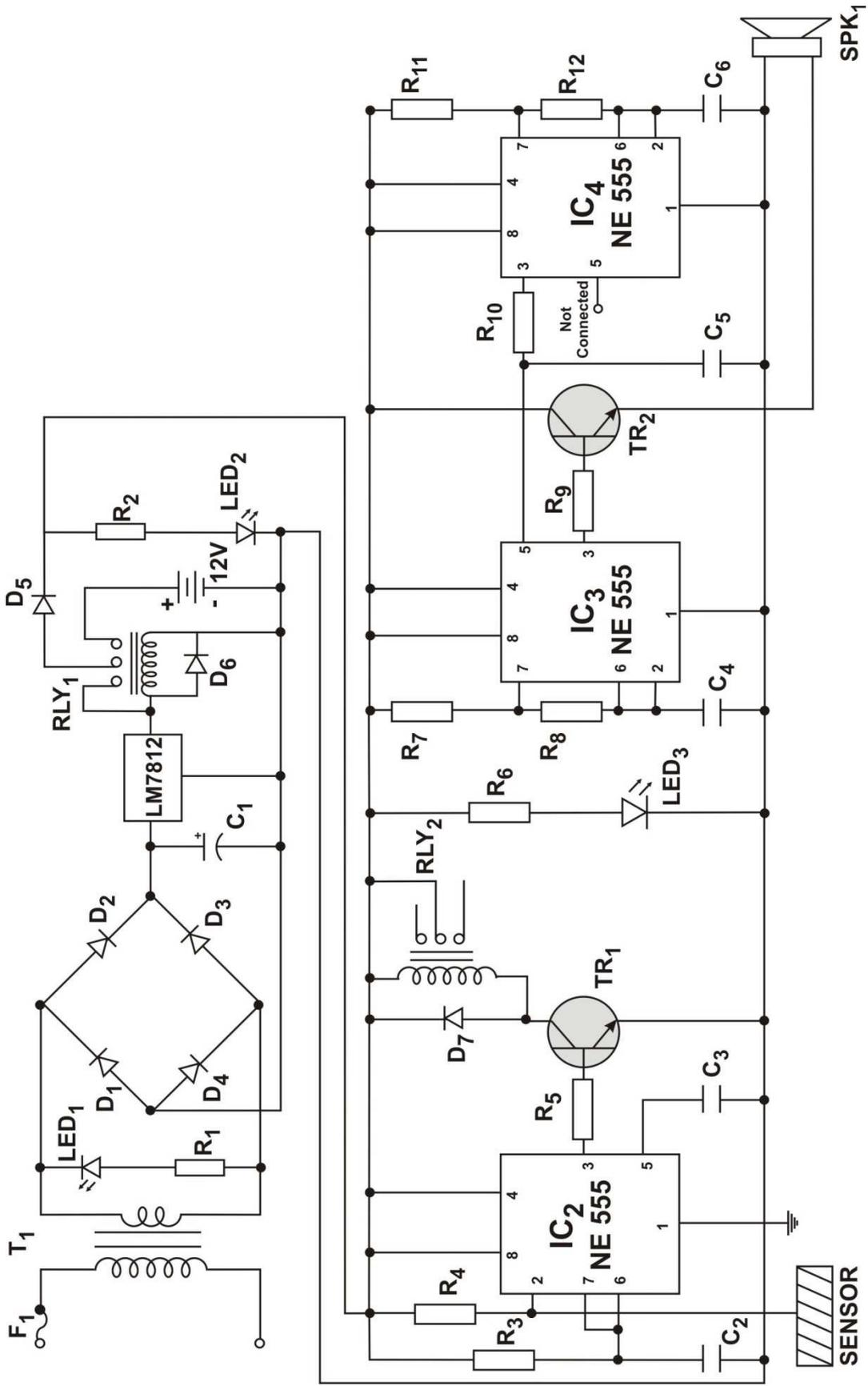

Fig. 6: Circuit diagram of a simple and reliable touch activated security system

Table 1: Components List

| SYMBOLS | COMPONENTS | RATING |
|---|---|---|
| T1 | Transformer | 240/18V,9V |
| F1 | Fuse | 1A |
| SPK1 | Speaker | 5 Watts, 8Ω |
| IC1 | Linear Power Regulator | LM7812 |
| LED1, LED2, LED3 | Light Emitting Diode | NTE 3010 |
| RLY1, RLY2 | Relay Switch | 12V, 400Ω |
| D1, D2, D3, D4, D5, D6, D7 | Diodes | 1N4001 |
| C1 | Capacitor | 2200μF, 25V |
| C2, C5, C6 | Capacitor | 47μF, 25V |
| C3, C4 | Capacitor | 0.01μF |
| TR1 | Transistor | SC1815 |
| TR2 | Transistor | TIP31 |
| R1 | Resistor | 470Ω |
| R2 | Resistor | 980Ω |
| R3 | Resistor | 220KΩ |
| R4 | Resistor | 10.8MΩ |
| R5 | Resistor | 4.7KΩ |
| R6, R12 | Resistor | 1KΩ |
| R7, R8 | Resistor | 100KΩ |
| R9 | Resistor | 300KΩ |
| R10 | Resistor | 2.2KΩ |
| R11 | Resistor | 22KΩ |

### 3.2.1.   Monostable Operation of a Timing Circuit

The monostable mode of operation is as also known as one short operation of timing circuit. When a negative-going trigger pulse is supplied to the trigger input, the threshold on the lower comparator is exceeded, and this as well triggers the output circuit. The basic operation of the circuit, and the duration of the output pulse in seconds is approximately equal to:

$$T = 1.1 \times R \times C \text{ (in seconds)} \tag{1}$$

### 3.2.2.   Astable Operation of a Timing Circuit

In an Astable mode both the trigger and threshold inputs (pin 2 and 6) to the two comparators are connected together and to the external capacitor. The frequency of operation of the Astable circuit is dependent upon the values of $R_a$, $R_b$ and C. the period can be calculated with the formula:

$$T = 0.693 \times C \text{ } (R_1) \tag{2}$$

The time difference between the ON time and the OFF time is mainly dependent on the value of $R_a$ and $R_b$. The ratio of the time duration when the output pulse is high to the total period is known as the duty-cycle and is represented as:

$$D = \frac{t_1}{t} \frac{(R_a + R_b)}{(R_a + 2R_b)} \tag{3}$$

$$t_1 = 0.693(R_a + R_b)C \tag{4}$$

$$t_2 = 0.693 x R_b x C \tag{5}$$

### 3.3. System Design Analysis

### 3.3.1. The Power Stage

A 240/18v transformer was chosen because its rating is capable of meeting the current demand of the circuit and it is protected by the 1A fuse against excess current. The limiting resistor (R1) for the LED1 was calculated as:

$$R_1 = \frac{Voltage drop}{LED current} \tag{6}$$

$$R_1 = \frac{(V_{CC} - V_{LED})}{I_{LED}} \tag{7}$$

Where $V_{CC}$ = supply voltage = 18V, $V_{LED}$ = 2.2, $I_{LED}$ = maximum allowable current across the LED = 35mA.

$$R_1 = \frac{(18 - 2.2)V}{35mA} = \frac{15.8V}{0.035A}$$

$$R_1 = 451.43\Omega$$

The preferred resistor value closest to $451.43\,\Omega$ is $470\,\Omega$. Therefore $470\,\Omega$ were adopted in the design.

Current drawn by $LED_1 = \frac{18V}{470\Omega} = 0.038MA$

The peak inverse voltage (PIV) obtainable at the secondary terminal transformer is twice the terminal voltage $V_s$ [13].

That is: PIV = 2 x Vs = 2 x 18 = 36V.

At the full bridge rectifier circuit IN4001 diode was used because its PIV which is 50V is greater than the PIV of the secondary of the secondary terminal which is 36V[13]. This was done to avoid damage to the diodes in case reverse operation occurs. The value of the filter capacitor $C_1$ was obtained as:

$$C = \frac{1}{4\sqrt{3}fyR} \tag{8}$$

(For full wave rectifier circuits)[13], where: f = frequency of ripple voltage = 50Hz

y = ripple factor = 5% = 0.05

R = resistance of the regulator $= \dfrac{V}{I}$

(9)

V = Constant output voltage from the regulator = 12v, I = Constant output current from the regulator = 500mA = 0.5A

$$R = \frac{12}{0.5} = 24\Omega$$

$$C = \frac{1}{4x\sqrt{3x24x50x0.05}} , \quad C = \frac{1}{415.692}$$

$$C = 2.4056x10^{-6} \, LEDcurrent , \quad C = 2.405.6x\mu F$$

A 2200 $\mu F$ was used in the design because it is the closest value of a standard capacitor of $2.405.6x\mu F$. The current limiting resistor for LED2 is calculated as shown below:

$$R_2 = \frac{Voltagedrop}{LEDcurrent}$$

$$R_2 = \frac{Vcc - V(LED)}{I(LED)}$$

Where $V_{CC}$ = supply voltage = 12V, $V_{LED}$ = 2.2, I (LED) = 0.01A (Chosen to limit the amount of current consumed by the LED)

$$R_2 = \frac{(12 - 2.2)V}{0.01A}, \quad R_2 = \frac{9.8V}{0.01A}$$

$$R_2 = 980\Omega$$

The preferred resistor value closest to $980\Omega$ is $1K\Omega$. Therefore $1K\Omega$ resistor was adopted as $R_2$ in the design.

Current drawn by LED2 $= \dfrac{12V}{1000\Omega} = 0.012A = 12mA$

### 3.3.2. The Trigger Stage

The timeout period (T) and the frequency (f) were determined by the values of $R_4$ and $C_2$ as follows:

$$T = 1.1(R_3 x C_2)$$

(10)

But R3 = 220KΩ = 220 x $10^3\Omega$, C2 = 47μF = 47 x $10^{-6}$F

$$T = 1.1(220x10^3 220x47x10^{-6}) \sec s = (1.1x10.34) \sec s$$

$$= 11.374 \sec s \approx 11 \sec s$$

$$f = \frac{1}{T} = \frac{1}{11.374} = 0.09Hz$$

The values of R$_4$ and C$_2$ were chosen in such a way that they can produce an approximate period of 11secs delay. The basis resistor R$_5$ for transistor Tr1 was chosen as a result of the following calculations:

$$R_6 = \frac{Vcc - V_{BE}}{I_B}$$

(11)

Where V$_{CC}$ = supply voltage = 12V, V$_{BE}$ = Base emitter voltage = 0.6(see appendix III)

I$_B$ = Base current

I$_B$ = I$_C$/gain Supply voltage, Gain = 25 (h$_{FE}$)

I$_C$ = collector current = maximum Relay current = $\dfrac{Supply Voltage}{Coil\,\mathrm{Re}\,sis\tan ce} = \dfrac{12}{400} = 0.03A$

$$\Rightarrow I_B = \frac{0.03}{25} = 0.0012A$$

To ensure that the current is sufficient to drive the transistor into saturation, the quantity of the current is doubled i.e.

$$I_B = 0.0012 x2 = 0.024A$$

$$R_5 = \frac{12 - 0.6}{0.0024} = \frac{11.4}{0.0024}$$

$$47050\Omega \approx 4.7\Omega K$$

A $4.7\Omega K$ resistor was chosen to serve as the base resistor (R6) to the transistor because it is the closet value of standard resistor value to $47050\,\Omega$.

### 3.3.3. The Alarm/Amplifier Stage

Design calculation for this unit was done in three stages.

**Stage 1:** This is the high frequency oscillator stage. The period (T$_H$) and frequency (f$_H$) for this stage were calculated as follows:

$$T_H = t_1 + t_2$$

(12)

t$_1$ = 0.693 x C$_4$ (R$_7$ + R$_8$)

But C4 = 0.01µF = 0.01 x 10$^{-6}$F

R$_7$ = R$_8$ = 100KΩ =100 x 10$^3$Ω

Therefore, t$_1$ = 0.693 x 0.01 x 10$^{-6}$(100 x 10$^3$ + 100 x 10$^3$)

= 0.693 x 0.01 x 10$^{-6}$ x 200 x 10$^3$ = 1.386 x 10$^{-3}$secs

t$_2$= 0.693 x 100 x 10$^3$ x 0.01 x 10$^{-}$

t$_2$= 0.693 x 10$^{-3}$

∴ T$_H$ = t$_1$ + t$_2$ =1.386 x 10$^3$ + 0.693x 10$^{-3}$ = 2.079 x 10$^{-3}$secs = 2.079msecs

$$F_H = \frac{1}{T_H} = \frac{1}{2.097 x10^{-3}} = 481\text{Hz}$$

Duty cycle $= \dfrac{t_1}{T_H} = \dfrac{1.38 x 10 - 3}{2.097 x 10^{-3}} x 100\%$ $= 66.95\%$

The value of $C_4$, $R_7$ and $R_9$ were manipulated in order to get the desired frequency that will modulate the low frequency oscillator ($IC_2$) to give the desired tone.

**Stage 2:** This is the low frequency oscillator stage. The period ($T_L$) and frequency ($F_L$) for this were calculated as follows:

$$T_L = t_1 + t_2 \tag{13}$$

$t_1 = 0.693$ x $C_6$ ($R_{11} + R_{12}$)

But $C_6 = 47\mu F = 47$ x $10^{-6}F$

$R_{11} = 1K\Omega = 1$ x $10^3$ $\Omega$

Therefore, $t_1 = 0.693$ x $47$ x $10^{-6}$ ($1$ x $10^3 + 22$ x $10^3$) $= 0.693$ x $47$ x $10^{-6}$ x $23$ x $10^3$

$= 10741.003$x $10^{-3}$secs $= 1.041$secs

$t_2 = 0.693$ x $C_6$ x$R_{12}$

$t_2 = 0.693$ x $47$ x $10^{-6}$ $22$ x $10^3$

$t_2 = 995.74$ x $10^{-3}$

$= 0.0996$secs

$\therefore T_L = t_1 + t_2 = 1.041 + 0.996 = 2.037$secs

$F_L = \dfrac{1}{T_L} = \dfrac{1}{2.037} = 0.491$Hz

Duty cycle $= \dfrac{t_1}{T_L} = \dfrac{1.0411}{2.037}$x$100\%$ $= 51.1\%$

The values of $C_6$, $R_{11}$ and $R_{12}$ were manipulated in order to get the desired frequency that will modulate the high frequency oscillator ($IC_3$) to give the desired tone.

**Stage 3:** This is the audio amplifier stage. It is this stage that gives the final power output to the speaker. The actual power output by the transistor (Tr2) was calculated as follows:

$$\text{Power output} = I_E \text{ x } V_{CC} \tag{14}$$

Where $V_{CC}$ = supply voltage = 12v, $I_E$ = Emitter current

But $I_E = (1 + \text{gain})$ x $I_B$

Gain = 100 ($h_{FE}$)

$I_B = \dfrac{12 - 0.6}{300} = \dfrac{11.4}{300} = 0.038A$

$I_E = (1 + 100)$ x $0.038 = 0.418A$, and Substituting back into to Eqn. (14),

$\Rightarrow P_{OUT} = 0.418x12 = 5.016$ Watts

$\approx 5$Watts

This means that the amplifier stage of the alarm unit (Tr2) gives an output of 5Watts. Therefore a 5Watts, 8$\Omega$ speaker was chosen at the output stage for maximum power transfer.

# 4. RESULTS, TESTING AND DISCUSSIONS

## 4.1. Construction and Casing

This section described the steps taken in the verification of calculated results through the real time implementation and measurements. The construction of the system is in 2 stages; the soldering of the components and the coupling of the entire system to the casing. The power supply stage was first soldered, and then the transmitter and receiver stage and all the other stages were soldered. The circuit was soldered in a number of patterns that is, stage by stage. Each stage was tested using the multi-meter to make sure it is working properly before the next stage is done. This helps to detect mistakes and faults easily. The soldering of the circuit was done on a 10cm by 24cm Vero-board. The second stage of the system construction is the casing of the soldered circuit. Casing refers to the outer covering or something that serves as a container or covering. For the purpose of this system, the material used for the casing was a metal sheet. Proper dimensioning of the casing was marked out to give the desired shape based on the size of the constructed project work on Vero-board. After fabricating the casing, finishing was done on it which involves smoothening with file and painting the casing to give aesthetic values to the system. The final casing that was obtained at the end of the construction is shown in Fig. 7 below.

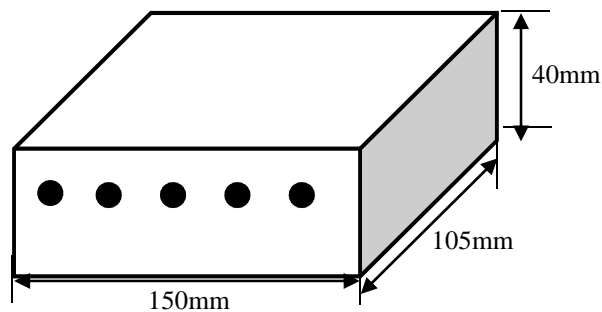

Fig. 7: System Main Casing

## 4.2. Testing and Results

In testing the designed and constructed system, four basic steps were taken. These steps are sequentially listed below as:

**Step 1:** To ensure that all the components to be used are functionally operating, they were first tested with a digital multi meter and failed ones replaced before finally soldering them on the veroboard.

**Step 2:** To ensure that there was no breakage in the circuit path on the veroboard, immediately after soldering on veroboard, the circuit path was tested using the Digital Multi-meter. This was done to also ensure continuity of circuit on the veroboard.

**Step 3:** Using Circuit Maker 6 (Student Edition), the circuit was simulated. The result obtained from the simulation closely corresponds to the desired result, with only some slight variations.

**Step 4:** The period of time for the alarm sound (Time out period) was manually tested. This was achieved using Digital Stop Watch and the result obtained was found to be 10.60 seconds. The value obtained from the manual testing closely agrees with that obtained in the design specifications i.e. 11.37 seconds.

### 4.3. Discussion of Results

The main reason for testing all the components before they were finally soldered on the veroboard is to avoid the painstaking effort it will take to dis-solder faulty components at the end of the day. From the continuity test carried out on the veroboard to check the circuit path, it was discovered that the circuit was in a perfect working condition as continuity was ensured. Simulation of the circuit design was also done as mentioned earlier, with the sole objective of comparing the results obtained from design calculations to that obtained from simulation. The two results when compared closely correspond with only a very slight discrepancy in values.

### 4.4. Problem Encountered

It should be noted at this juncture, that the realization of the final system work was not without problems. The various problems encountered during the design and implementation stage are highlighted as: (1) some basic components to be used for the system were not within reach as it was not available in town. (2) Some measuring instruments that would have being used for detailed analysis of the circuit (i.e. Oscilloscope, Transistor Tester) were limited for use. Simulation software was instead used for the circuit analysis. (3) The biggest problem encountered was at the implementation stage as the circuit was triggering itself without touch. The sensitivity of the circuit was reduced by reducing resistance R2 to solve this problem.

## 5. CONCLUSIONS

It can be concluded that the sole aim of carrying out the design, analysis and implementation of *a simple and reliable touch sensitive security system* was achieved, in that the aim was to develop a cheap, affordable, reliable and efficient security system, which was successfully realized at the end of the design process. One factor that accounts for the cheapness of the product was the proper choice of components used. The ones that were readily available were used, while a close substitute was found for those that were not readily available. The reliability of the entire alarm system was considered by the integration of an automatic change over switch into the power supply unit such that the A.C mains supply and the battery are cold redundant. Thus, this guarantees constant supply of power to the main circuit. The efficiency of the entire system was put into consideration by the use of transistor in the common collector mode to couple the output of the circuit to the speaker. The system was tested and found to be working to specifications and predictions. Summarily, a cheap and reliable way of checking the activities of burglars and intruders has been successfully developed, which is the aim of the research. We can conclusively say therefore, that the benefits of having this burglar alarm system cannot be overemphasized. In future, we shall find a way of improving the system by interfacing the alarm system with the microcomputer to boost the effectiveness of the entire system or integrating a digital door lock.

### ACKNOWLEDGEMENTS

The authors would like to thank Col. Muhammed Sani Bello (RTD), OON, Vice Chairman of MTN Nigeria Communications Limited for supporting the research.

**Author**

**Engr. Adamu Murtala Zungeru** received his B.Eng. degree in Electrical and Computer Engineering from the Federal University of Technology (FUT) Minna, Nigeria in 2004, and M.Sc. degree in Electronic and Telecommunication Engineering from the Ahmadu Bello University (ABU) Zaria, Nigeria in 2009. He is a Lecturer Two (LII) at the Federal University of Technology Minna, Nigeria in 2005-till date. He is a registered Engineer with the Council for the Regulation of Engineering in Nigeria (COREN), Member of the Institute of Electrical and Electronics Engineers (IEEE), and a professional Member of the Association of Computing Machinery (ACM). He is currently a PhD candidate in the department of Electrical and Electronic Engineering at the University of Nottingham. His research interests are in the fields of automation, Security, swarm intelligence, routing, wireless sensor networks, energy harvesting, and energy management.

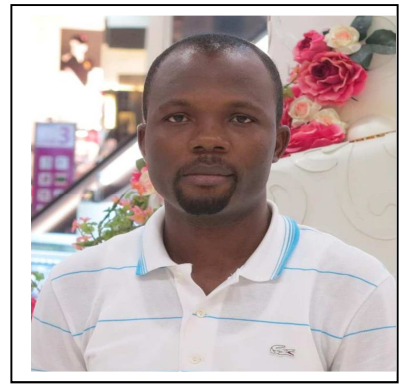